\documentclass{PoS}

\title{Dark matter from one-flavor SU(2) gauge theory}

\ShortTitle{Dark matter from one-flavor SU(2) gauge theory}

\author{Anthony Francis, Renwick James Hudspith, \speaker{Randy Lewis}, Sean Tulin \\
      Department of Physics and Astronomy, York University, Toronto, Ontario, M3J 1P3, Canada \\
      E-mail: \email{afrancis.heplat@googlemail.com}, \email{renwick.james.hudspith@googlemail.com}, \email{randy.lewis@yorku.ca}, \email{stulin@yorku.ca}}

\abstract{SU(2) gauge theory with a single fermion in the fundamental representation is a minimal non-Abelian candidate for the dark matter sector, which is presently missing from the standard model. Having only a single flavor provides a natural mechanism for stabilizing dark matter on cosmological timescales. Preliminary lattice results are presented and discussed in the context of dark matter phenomenology.}

\FullConference{34th annual International Symposium on Lattice Field Theory\\
		 24-30 July 2016\\
		 University of Southampton, UK}

\begin{document}

\section{Motivation}

As the standard model of particle physics does not explain
all of the mass in galaxies and galaxy clusters, the existence of dark
matter is inferred \cite{Kapteyn:1922zz,Zwicky:1933gu}.
The dark sector could be a non-Abelian gauge theory, and the lattice approach
is a valuable tool for studying dark matter candidates of this type
\cite{Lewis:2011zb,Hietanen:2012sz,Appelquist:2013ms,Hietanen:2013fya,Appelquist:2014jch,Hietanen:2014xca,Detmold:2014qqa,Detmold:2014kba,DeGrand:2015lna,Appelquist:2015zfa,Drach:2015epq,Arthur:2016ozw}.
For a recent review, see \cite{Kribs:2016cew}.

In the present work, we consider a minimal non-Abelian gauge theory
for dark matter, namely
SU(2) with a single fermion in the fundamental representation.
It has similarities to QCD but also interesting differences, as will be
discussed below.  Of particular relevance to dark matter phenomenology is
the fact that this simple theory can stabilize its dark matter candidate
in a natural way, in contrast to SU(2) with two or more fundamental
fermions for example, as discussed in section \ref{SM}.

\section{Basic theory}

The Lagrangian for SU(2) gauge theory with one Dirac flavor can be written as
\begin{equation}\label{Lagrangian}
{\cal L} = -\frac{1}{4}F_{\mu\nu}^aF^{a\,\mu\nu}+i\bar{Q}\gamma^\mu D_\mu Q + mQ^TCEQ
\end{equation}
where
\begin{equation}\label{QandE}
Q = \left(\begin{array}{c} \chi_L \\ C\bar\chi_R^T \end{array}\right),~~~
E = \left(\begin{array}{cc} 0 & 1 \\ -1 & 0 \end{array}\right)
\end{equation}
and $C$ is the charge conjugation matrix.
The dark fermion field $Q$ has no standard model quantum numbers, but in a
later section we will describe its interaction with the standard model through
couplings to the Higgs sector.

The Lagrangian in eq.~(\ref{Lagrangian}) has an unbroken global SU(2), namely
\begin{equation}
Q\to e^{i\sum_{i=1}^3T_i\alpha_i}Q
\end{equation}
which is a generalization of baryon number.
In a theory with $N_f>1$ fermion flavors, the global SU(2) would be SU($2N_f$).
For lectures describing multi-flavor SU(2) gauge theory, see \cite{vonSmekal:2012vx}.

For $m=0$, the Lagrangian in eq.~(\ref{Lagrangian}) also has an unbroken
(but anomalous) global U(1), namely
\begin{equation}
Q\to e^{i\beta}Q
\end{equation}
like the axial U(1) in QCD.
We expect this U(1) symmetry to be broken dynamically by a mass-like vacuum expectation value.

The particle spectrum of one-flavor SU(2) will include mesons, baryons and glueballs, all of which must appear as multiplets of the global SU(2).
The simplest creation operators for mesons have the form $\bar Q\Gamma Q$.
The simplest creation operators for baryons have the form $Q^TC\Gamma EQ$
where $C$ is the charge conjugation matrix and $E$ was defined in eq.~(\ref{QandE}).
One important example is
$\bar Q\gamma_5Q$ which is a singlet denoted by $\eta$.
Another important example is
$\bar Q\gamma_\mu Q$ which is one entry in a triplet where the other two
entries are the baryon and anti-baryon; we will name this triplet $\rho^{\pm,0}$.

\section{Preliminary lattice explorations}

Our lattice studies use a standard discretization of eq.~(\ref{Lagrangian}):
the plaquette gauge action and the Wilson fermion action.
This is achieved efficiently by using an up-to-date version of the HiRep code \cite{DelDebbio:2008zf}.
The lattice action contains two parameters: the inverse gauge coupling $\beta$
and the bare fermion mass $m_{\rm bare}$.
We have generated two ensembles with the $\beta$ values tuned to give
comparable physical volumes, and for each ensemble
we calculated propagators at the sea fermion mass as well as
propagators at additional valence fermion masses.
These partially quenched results are used to study
fermion mass dependences.  Further details of our ensembles are displayed in
Table~\ref{latticeparameters}.

\begin{table}[t]
\begin{center}
\begin{tabular}{c|cc}
\hline
$\beta$ & 2.2 & 2.309 \\
lattice dimensions & $20^3\times56$ & $28^3\times56$ \\
number of configurations & 2000 & 1540 \\
acceptance & 73\% & 74\% \\
unitary $am_{\rm bare}$ & -0.865 & -0.76 \\
partially quenched $am_{\rm bare}$ & -0.845, -0.855 & -0.74, -0.75 \\
average plaquette & 0.5989 & 0.6255 \\
$aw_0$ & 1.430(5) & 1.956(7) \\
$m_VL$ & 9.0(2) & 8.8(2) \\
\hline
\end{tabular}
\caption{Lattice parameters for each of our two lattice ensembles.}\label{latticeparameters}
\end{center}
\end{table}

\begin{figure}[b]
\begin{center}
\includegraphics[width=10cm,trim=0 40 0 0,clip=true]{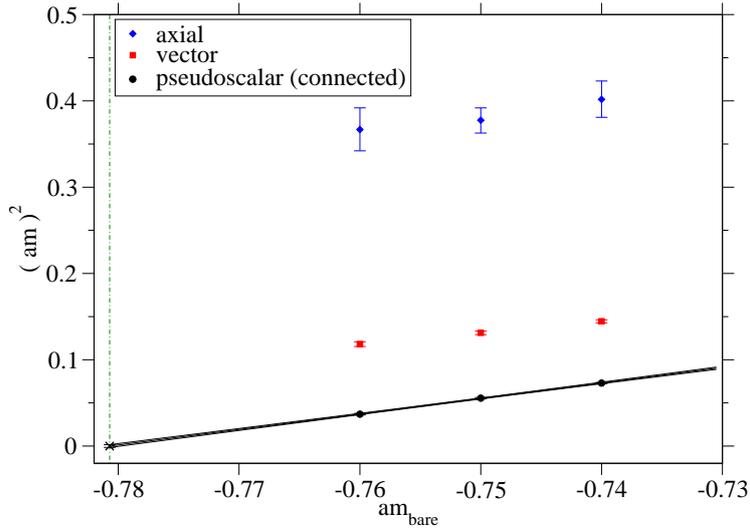}
\vspace{-5mm}
\end{center}
\caption{Three hadron masses (squared) as a function of the bare fermion mass at $\beta=2.309$.}\label{spectrum1}
\end{figure}

The hadron mass spectrum at the smaller lattice spacing, neglecting disconnected
diagrams, is shown in figure~\ref{spectrum1}.
This absence of disconnected diagrams removes the anomaly
contribution to $m_\eta^2$, resulting in a linear dependence on the bare fermion
mass as if it were a Goldstone boson.
When disconnected diagrams are included, the $\eta$ should not be an
authentic Goldstone boson due to the anomaly.
Note that disconnected diagrams have recently been studied in the two-flavor theory \cite{Arthur:2016ozw}.

\begin{figure}
\begin{center}
\includegraphics[width=10cm,trim=0 40 0 0,clip=true]{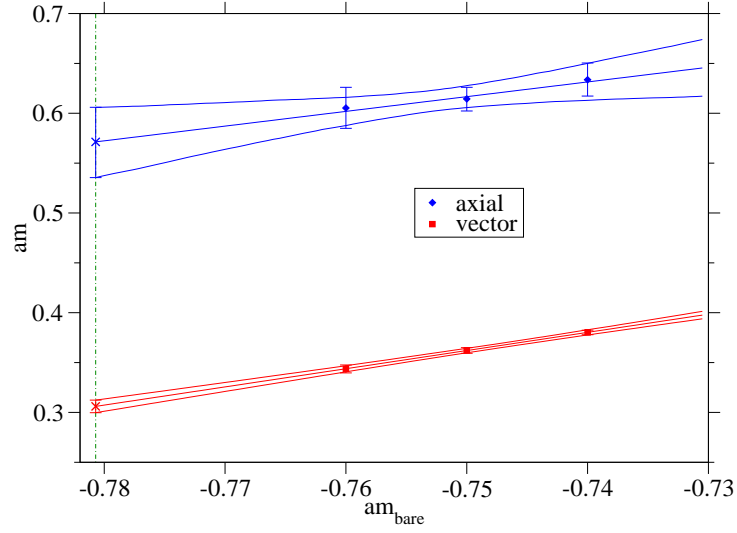}
\vspace{-5mm}
\end{center}
\caption{Linear extrapolations for the vector and axial vector meson masses as a function of the bare fermion mass at $\beta=2.309$.}\label{spectrum2}
\end{figure}

As in QCD, the axial meson is heavier than the vector which is heavier than the
pseudoscalar.  Recall that the lightest baryon and anti-baryon are exactly
degenerate with the vector meson due to the global SU(2).
Linear extrapolations of the vector and axial masses are displayed in
figure~\ref{spectrum2}.
The vertical dashed line is where the connected part of $m_\eta=0$, and in that limit the axial meson mass is approximately twice as large as the vector meson mass.

\begin{figure}[b]
\includegraphics[height=55mm,trim=30 40 0 0,clip=true]{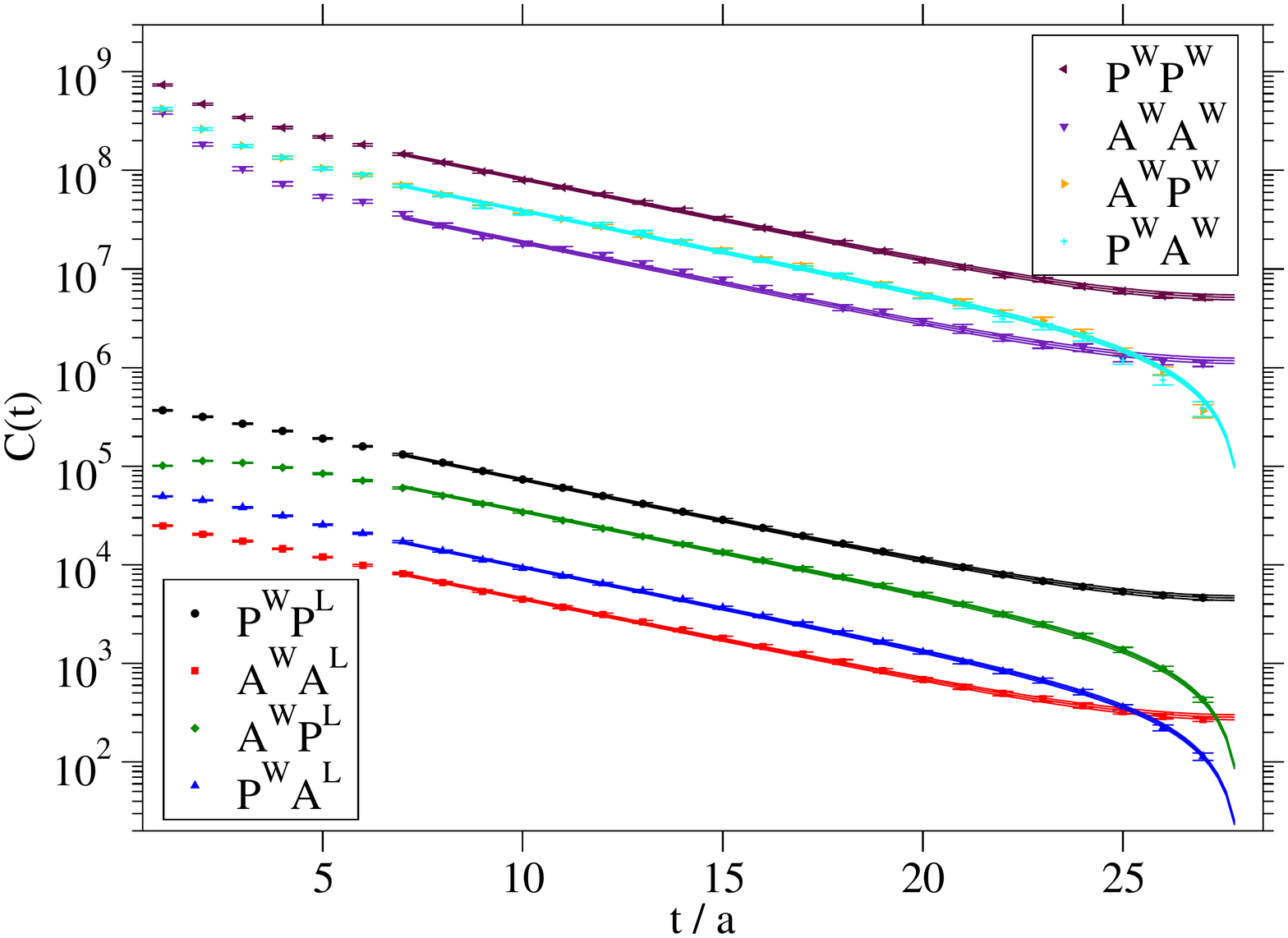}
\includegraphics[height=55mm,trim=0 40 0 0,clip=true]{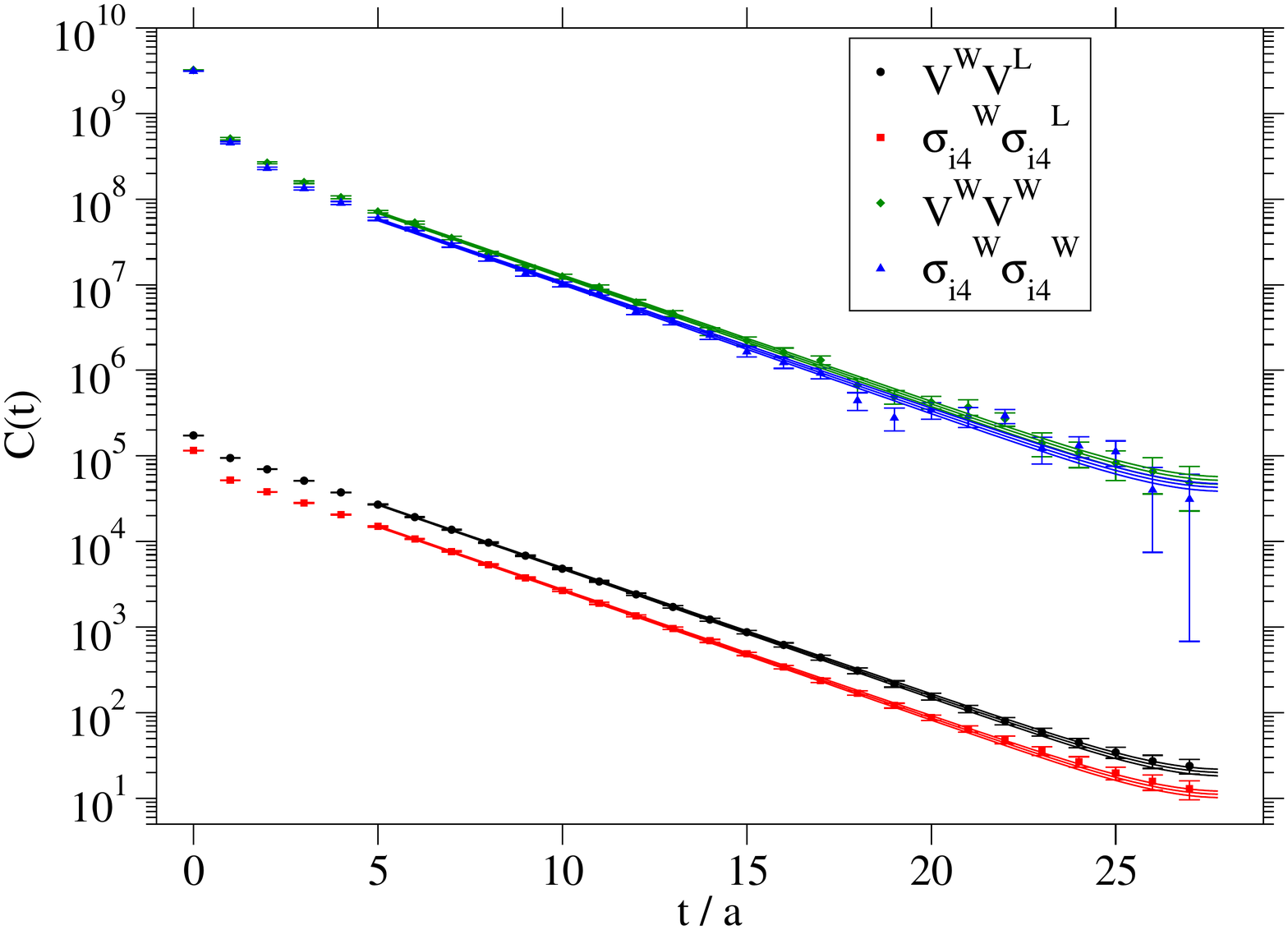}
\caption{Correlation functions at $\beta=2.309$ and $am_{\rm bare}=-0.76$
for connected pseudoscalar (left panel) and vector (right panel).
L denotes a local point sink and W is a Coulomb gauge fixed wall. P, A, V and $\sigma_{i4}$ are conventional Dirac bilinears.}\label{correlators}
\end{figure}

Some of the raw correlation functions that generated our hadron mass results
are presented in figure~\ref{correlators}.
We use Coulomb gauge fixed wall sources and perform contractions
with a wall sink or a point sink.
The resulting correlation functions have small statistical uncertainties and
are dominated by the ground state, so one-state fits
give precise hadron masses.

\section{A large $N_c$ limit}

The standard limit of SU(2)$\to$SU(3)$\to$SU(4)$\to\ldots$ is not useful
in the present context because the global SU(2) symmetry that shapes the
spectrum of SU(2) gauge theory does not exist for any $N_c>2$.
However, SU(2) gauge theory is also Sp(2) gauge theory, and the global SU(2)
symmetry is indeed present for any Sp($N_c$) gauge theory with even $N_c$.
Therefore the useful limit is Sp(2)$\to$Sp(4)$\to$Sp(6)$\to\ldots$.
The $\eta$ becomes massless as $N_c\to\infty$ and $m_Q\to0$.
In this double limit, the global U(1) is only broken
dynamically and the $\eta$ is its Goldstone boson.

It might seem surprising that the global SU(2) can really remain unbroken
as $N_c\to\infty$.  A baryon with fermion content
$X=\sum_{i,j,\cdots k=1}^{N_c}Q_iQ_j\ldots Q_k$
cannot be inside the same multiplet as a meson with fermion content
$M=\sum_{i=1}^{N_c}\bar Q_i Q_i$
since they clearly contain differing numbers of fermions.
The resolution is that $X$ is actually a collection of baryons; an
individual baryon is a two-fermion object in Sp($N_c$) gauge theory:
$B=\sum_{i=1}^{N_c}\sum_{j=1}^{N_c}Q_iE_{ij}Q_j$.
In this way, the global SU(2) remains unbroken.
Baryons $B$ and mesons $M$ remain degenerate.

\section{Coupling to the standard model}\label{SM}

Dark matter appears to have no direct couplings to the standard model
gauge theories, but there can be couplings to the standard model Higgs.
In the one-flavor SU(2) theory, the leading interaction is a dimension 5 term:
\begin{equation}
\delta{\cal L} \propto \frac{1}{\Lambda} \bar{Q}\gamma_5Qh^\dagger h
\end{equation}
which will couple to the dark $\eta$ meson. $\Lambda$ denotes the scale of
new BSM physics.
For the dark $\rho$ triplet, the leading Higgs interaction is dimension 6:
\begin{equation}
\delta{\cal L} \propto \frac{1}{\Lambda^2} \bar{Q}\gamma_\mu Qh^\dagger\nabla^\mu h.
\end{equation}
Our dark matter candidate is the $\rho$, which can be stable for the life of the universe: recall for example that standard model proton decay is also a
dimension 6 interaction.
The dark $\eta$ will not be so long lived.
It should be noted that $\rho$ stability is a consequence of having a single
fermion flavor: with multiple flavors there are dimension 5 operators for
$\rho$ decay.

Without assuming parity conservation in the dark sector, the
mass terms for the dark fermion are
\begin{equation}
\delta{\cal L} = -m_4\cos\theta_4\bar{Q}Q-m_4\sin\theta_4\bar{Q}i\gamma_5Q
-\frac{v^2}{\Lambda}\cos\theta_5\bar{Q}Q\left(1+\frac{h}{v}\right)^2
-\frac{v^2}{\Lambda}\sin\theta_5\bar{Q}i\gamma_5Q\left(1+\frac{h}{v}\right)^2
\label{massterms}
\end{equation}
where $v=246$ GeV is the electroweak scale.
Notice that the dark fermion gets mass from 2 sources:
from BSM at dimension 4 and through the Higgs at dimension 5.

The mass terms can be written more compactly as
$\delta{\cal L} = m\bar Q_{\rm tw}Q_{\rm tw}$ where
the twisted field is $Q_{\rm tw} \equiv e^{i\gamma_5\alpha/2}Q$.
Our use of the standard Wilson lattice action corresponds to implicitly
untwisting the original mass terms in eq.~(\ref{massterms}).
Vector and axial hadrons are invariant under $Q\to Q_{\rm tw}$.

\begin{figure}[t]
\begin{center}
\includegraphics[width=10cm,trim=0 30 0 80,clip=true]{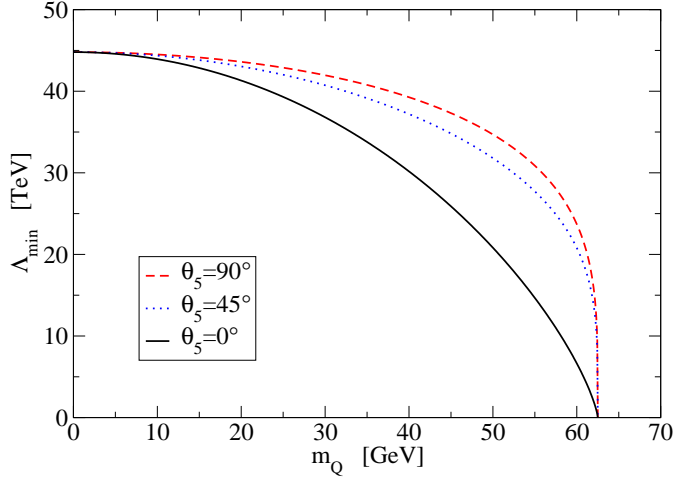}
\vspace{-5mm}
\end{center}
\caption{The minimum BSM scale as a function of dark fermion mass, for various choices of $\theta_5$ in eq.~(\protect\ref{massterms}).}\label{invisibleHiggs}
\end{figure}

The current experimental bound on invisible Higgs decays,
$\Gamma(h\to Q\bar{Q})<1.2$ MeV, provides a lower bound on the BSM scale in our dark
sector, as shown in figure~\ref{invisibleHiggs}.
The typical scale is seen to be tens of TeV.

\begin{figure}[t]
\begin{center}
\includegraphics[width=10cm,trim=0 40 0 0,clip=true]{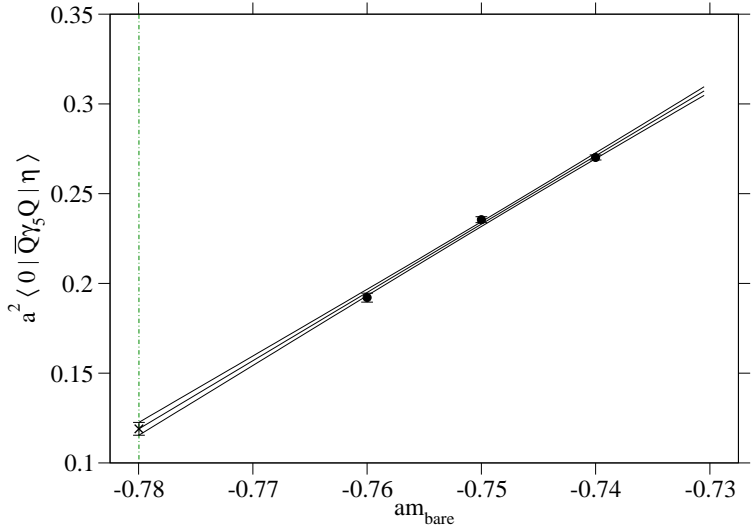}
\vspace{-5mm}
\end{center}
\caption{The connected part of the $\eta$ decay amplitude at $\beta=2.309$.}\label{etadecay}
\end{figure}

The only decays from the dark sector into standard model particles are
through Higgs bosons.
With the $\rho$ as our, essentially stable, dark matter candidate,
an additional constraint comes from big bang nucleosynthesis,
which requires the $\eta$ lifetime to be less than about 1 second \cite{Gamow:1946eb}.
This means that a lattice determination of
$\langle0|\bar Q\gamma_5Q|\eta\rangle$ would provide a bound on
$\frac{\sin\theta_5}{\Lambda}$.
A calculation for $m_\eta\ll m_h$ gives
\begin{equation}
\Gamma_\eta = \left|\langle0|\bar Q\gamma_5Q|\eta\rangle\right|^2\frac{m_\eta\sin^2\theta_5}{2\pi\Lambda^2m_h^4}\sum_{f\in{\sf SM}}m_f^2\left(1-\frac{4m_f^2}{m_\eta^2}\right)^{3/2}.
\end{equation}
A complete lattice calculation must include the disconnected diagrams,
but a proof of concept is given in figure~\ref{etadecay},
which shows the matrix element obtained from a lattice calculation
of the connected part only.

\section{Outlook}

One-flavor SU(2) gauge theory provides a minimal non-Abelian dark sector.
It has a global SU(2) to stabilize dark matter (here called the $\rho$),
and no dark matter decay at dimension 5.
One Goldstone boson should emerge for Sp($N_c\to\infty$), and
disconnected diagrams must be computed to arrive at a complete lattice calculation.

One phenomenologically important topic left for future work is direct
detection, which entails $\rho$ scattering from a standard
model nucleon. Another is the relic density, which requires a lattice
calculation that can determine whether the $\rho$ is heavier or lighter than
the $\eta$.
In particular, $m_\rho>m_\eta$ could have a relic density governed by
$\rho\rho\to\eta\eta$ or asymmetric dark matter, whereas $m_\rho<m_\eta$
could be governed by $\rho\rho\rho\to\rho\rho$.

\acknowledgments

We are grateful to Claudio Pica for help in the initial stages of this work,
and for providing access to the HiRep code.
Calculations were performed on Compute Canada's GPC machine at SciNet.
This work was supported in part by NSERC of Canada.


\begin{thebibliography}{99}
\bibitem{Kapteyn:1922zz} 
  J.~C.~Kapteyn,
  %``First Attempt at a Theory of the Arrangement and Motion of the Sidereal System,''
  Astrophys.\ J.\  {\bf 55}, 302 (1922).
\bibitem{Zwicky:1933gu} 
  F.~Zwicky,
  %``Die Rotverschiebung von extragalaktischen Nebeln,''
  Helv.\ Phys.\ Acta {\bf 6}, 110 (1933).
\bibitem{Lewis:2011zb}
  R.~Lewis, C.~Pica and F.~Sannino,
  %``Light Asymmetric Dark Matter on the Lattice: SU(2) Technicolor with Two Fundamental Flavors,''
  Phys.\ Rev.\ D {\bf 85}, 014504 (2012)
  [arXiv:1109.3513].
\bibitem{Hietanen:2012sz}
  A.~Hietanen, C.~Pica, F.~Sannino and U.~I.~S\o ndergaard,
  %``Orthogonal Technicolor with Isotriplet Dark Matter on the Lattice,''
  Phys.\ Rev.\ D {\bf 87}, 034508 (2013)
  [arXiv:1211.5021].
\bibitem{Appelquist:2013ms}
  T.~Appelquist {\it et al.}  [LSD Collaboration],
  %``Lattice calculation of composite dark matter form factors,''
  Phys.\ Rev.\ D {\bf 88}, 014502 (2013)
  [arXiv:1301.1693].
\bibitem{Hietanen:2013fya} 
  A.~Hietanen, R.~Lewis, C.~Pica and F.~Sannino,
  %``Composite Goldstone Dark Matter: Experimental Predictions from the Lattice,''
  JHEP {\bf 1412}, 130 (2014)
  [arXiv:1308.4130].
\bibitem{Appelquist:2014jch}
  T.~Appelquist {\it et al.}  [LSD Collaboration],
  %``Composite bosonic baryon dark matter on the lattice: SU(4) baryon spectrum and the effective Higgs interaction,''
  Phys.\ Rev.\ D {\bf 89}, 094508 (2014)
  [arXiv:1402.6656].
\bibitem{Hietanen:2014xca}
  A.~Hietanen, R.~Lewis, C.~Pica and F.~Sannino,
  %``Fundamental Composite Higgs Dynamics on the Lattice: SU(2) with Two Flavors,''
  JHEP {\bf 1407}, 116 (2014)
  [arXiv:1404.2794].
\bibitem{Detmold:2014qqa} 
  W.~Detmold, M.~McCullough and A.~Pochinsky,
  %``Dark Nuclei I: Cosmology and Indirect Detection,''
  Phys.\ Rev.\ D {\bf 90}, 115013 (2014)
  [arXiv:1406.2276].
\bibitem{Detmold:2014kba} 
  W.~Detmold, M.~McCullough and A.~Pochinsky,
  %``Dark nuclei. II. Nuclear spectroscopy in two-color QCD,''
  Phys.\ Rev.\ D {\bf 90}, 114506 (2014)
  [arXiv:1406.4116]].
\bibitem{DeGrand:2015lna} 
  T.~DeGrand, Y.~Liu, E.~T.~Neil, Y.~Shamir and B.~Svetitsky,
  %``Spectroscopy of SU(4) gauge theory with two flavors of sextet fermions,''
  Phys.\ Rev.\ D {\bf 91}, 114502 (2015)
  [arXiv:1501.05665]].
\bibitem{Appelquist:2015zfa} 
  T.~Appelquist {\it et al.},
  %``Detecting Stealth Dark Matter Directly through Electromagnetic Polarizability,''
  Phys.\ Rev.\ Lett.\  {\bf 115}, 171803 (2015)
  [arXiv:1503.04205].
\bibitem{Drach:2015epq} 
  V.~Drach, A.~Hietanen, C.~Pica, J.~Rantaharju and F.~Sannino,
  %``Template Composite Dark Matter: $SU(2)$ gauge theory with 2 fundamental flavours,''
  PoS LATTICE {\bf 2015}, 234 (2016)
  [arXiv:1511.04370]].
\bibitem{Arthur:2016ozw} 
  R.~Arthur, V.~Drach, A.~Hietanen, C.~Pica and F.~Sannino,
  %``$SU(2)$ Gauge Theory with Two Fundamental Flavours: Scalar and Pseudoscalar Spectrum,''
  arXiv:1607.06654].
\bibitem{Kribs:2016cew} 
  G.~D.~Kribs and E.~T.~Neil,
  %``Review of strongly-coupled composite dark matter models and lattice simulations,''
  Int.\ J.\ Mod.\ Phys.\ A {\bf 31}, 1643004 (2016)
  [arXiv:1604.04627].
\bibitem{vonSmekal:2012vx} 
  L.~von Smekal,
  %``Universal Aspects of QCD-like Theories,''
  Nucl.\ Phys.\ Proc.\ Suppl.\  {\bf 228}, 179 (2012)
  [arXiv:1205.4205].
\bibitem{DelDebbio:2008zf} 
  L.~Del Debbio, A.~Patella and C.~Pica,
  %``Higher representations on the lattice: Numerical simulations. SU(2) with adjoint fermions,''
  Phys.\ Rev.\ D {\bf 81}, 094503 (2010)
  [arXiv:0805.2058]].
\bibitem{Gamow:1946eb} 
  G.~Gamow,
  %``Expanding universe and the origin of elements,''
  Phys.\ Rev.\  {\bf 70}, 572 (1946); Phys.\ Rev.\ {\bf 74}, 505 (1948).
\end{thebibliography}
\end{document}